\documentclass[runningheads]{llncs}
\usepackage[english]{babel}
\usepackage[T1]{fontenc}
\usepackage{makecell}
\usepackage{graphicx,verbatim}
\usepackage{amsfonts}
\begin{document}
%
%\title{Exploring the Limits of Foundation Models for Lung Nodule Segmentation}
%\title{Prompt Robustness and Automated Prompt Generation for Foundation Model Segmentation of Lung Nodules}
\title{What Matters is the Prompt: Prompt Sensitivity and Prompt Generation in Foundation Models for Lung Nodule Segmentation}


%\author{Jorfe F. Lazo\inst{1}\orcidID{0000-1111-2222-3333} \and
%Xixi Liu \inst{2,3}\orcidID{1111-2222-3333-4444} \and
%Third Author\inst{3}\orcidID{2222--3333-4444-5555}}
\author{Jorge F. Lazo\inst{1}\and
Xixi Liu \inst{1,2}\and
Andreas Hallqvist\inst{3, 4} \and
Mikael Johansson\inst{5} \and 
Åse Johnsson\inst{6,7} \and 
Jonas S. Andersson\inst{5} \and 
Jennifer Alvén\inst{1} \and 
Ida Häggström\inst{1, 8}}
\authorrunning{Jorge F. Lazo et al.}
\titlerunning{What Matters is the Prompt}
%\institute{Chalmers University of Technology, Gothenburg, Sweden \and
%Springer Heidelberg, Tiergartenstr. 17, 69121 Heidelberg, Germany
\institute{Department of Electrical Engineering, Chalmers University of Technology \and
Department of Computing, Imperial College London \and 
Department of Oncology, Sahlgrenska University Hospital \and 
Department of Oncology, Institute of Clinical Sciences, University of Gothenburg \and 
Department of Diagnostics and Intervention, Umeå University \and
Department of Radiology, Sahlgrenska University Hospital \and 
Department of Radiology, Institute of Clinical Sciences, University of Gothenburg \and 
Department of Medical Radiation Sciences, Institute of Clinical Sciences, University of Gothenburg
}

\maketitle              % typeset the header of the contribution
\begin{abstract}

Lung nodule segmentation in computed tomography is essential for extracting clinically relevant information for lung cancer assessment and treatment planning. 
Foundation models have shown notable segmentation capabilities, but state-of-the-art approaches often depend on input prompts, such as points or boxes, making their performance sensitive to prompt quality and placement.
Understanding the limitations and constraints of prompt-based foundation models is therefore essential for designing reliable medical image segmentation solutions. 
In this work, we investigate how prompt quality affects foundation models performance for lung nodule segmentation. 
We further propose a synthetic prompt-generation model to test if the dependence on manually provided prompts can be mitigated by generating synthetic prompts that can also improve segmentation performance. 
Perturbation experiments show that bounding box prompts generally outperform point prompts, while latest specialized medical imaging models achieve better performance than general purpose ones.
The proposed approach obtains a Dice coefficient of 0.85, suggesting that synthetic prompt generation as a promising strategy for lung nodule segmentation with foundation models.

%However, most of sota models typically rely on input prompts (e.g. seed points or boxes) to guide mask generation, making their performance dependent on the quality and placement of such prompts.
%In this work,  we analyze how the performance of different foundation models is affected by prompt quality in the task of lung nodule segmentation.  
%In addition, we propose a prompt-generation model to test if the dependence on manually provided prompts can be mitigated by generating synthetic prompts that can also improve segmentation performance. 
%Our results show that bounding box prompts lead to better segmentation performance than point prompts. 
%The automatic prompt generation model achieves a dice coefficient of 0.85 showing that the use of this method  can improve prompt-less segmentation. 

%Characterization
%How dependent are SAM/MedSAM on prompt quality?
%Quantification
%What level of prompt error can be tolerated?
%Mitigation
%Can automatically generated prompts satisfy these requirements?

% Authors must provide keywords and are not allowed to remove this Keyword section.

\end{abstract}
\section{Introduction}

Lung cancer is one of the deadliest and most common cancers in the world~\cite{zhu2024medical}. 
%Lung cancer in non-smokers is on the rise globally, now accounting for up to 20\% of new cases, particularly affecting women and younger adults.
Computed tomography (CT) is a widely used imaging technique for the preliminary diagnosis of lung cancer and the detection of lung nodules. Early detection is crucial to improving patient outcomes~\cite{de2020reduced}.
Nevertheless, identifying lung nodules is a complex and time-consuming task. 
%Lung nodules are small round oval-shaped structures in the lungs, typically less than 30 mm in diameter, and even medical experts struggle to differentiate between pulmonary nodules and surrounding organs as their visual appearance is similar. 
Accurate segmentation of lung nodules in CT scans is essential to provide critical information about their size, location, and extent for disease evaluation and treatment planning. However, obtaining precise segmentations of small anatomical structures, such as lung nodules, remains a challenge. 

With the rise of AI tools and methods, different models have been proposed to assist clinicians in the task of lung nodule detection and segmentation~\cite{tang2019nodulenet,setio2017validation}. 
Recently, foundation models have emerged as a promising paradigm for natural image segmentation, such as the Segment Anything (SAM)~\cite{Kirillov_2023_ICCV} model in natural images and MedSAM~\cite{ma2024segment} in the medical imaging domain.
%Image segmentation foundation models are large-scale pretrained vision models that learn generalizable representations for image segmentation from large and diverse datasets.
%
Unlike conventional segmentation models trained for a fixed set of object categories, segmentation foundation models are designed to perform segmentation across a wide variety of domains and object types, often through user prompts (e.g., points, bounding boxes, or text), with little or no task-specific fine-tuning. 
They serve as reusable foundations that can be adapted to numerous downstream segmentation tasks.

In the case of lung nodule segmentation, foundation models have mainly been explored as components within larger classification, detection and segmentation pipelines\cite{asha2024advanced,shaukat2024lung,swinburne2025foundational}. 
%rather than as fully autonomous segmentation systems. 
%Some studies have proposed a workflow based on SAM with transfer learning, where lung nodules are segmented from CT images using bounding-box prompts and the resulting representations are also used for nodule classification \cite{asha2024advanced}. 
%Similarly, Lung-CADex uses a MedSAM-based computer-aided detection module together with vision-language components for zero-shot nodule detection and malignancy classification, reducing the need for fully supervised training but still relying on a designed prompting and alignment strategy \cite{shaukat2024lung}. 
%Other studies further showed that SAM and MedSAM can be integrated with clinically mined data to obtain accurate lung cancer detection and segmentation models, suggesting that foundation models can reduce the annotation burden when combined with large-scale clinical data mining \cite{swinburne2025foundational}. 
%
More recent works have attempted to adapt foundation models more specifically to medical CT segmentation: StructSAM\cite{liu2026structsam} introduces structure-aware prompt adaptation to improve robustness for lung cancer lesion segmentation, while Probabilistic SAM extends SAM to produce multiple plausible masks and model annotation ambiguity, which is particularly relevant for lung nodules because boundaries can vary between experts \cite{ward2026probabilistic}. 
Other specialized models such as LNTransformer~\cite{ramezani2025lntransformer}use a fine-tuned variant of SAM to refine proposed segmentation masks given predicted boxes using deformable DETR. 
%
%An issue with foundation models is that they are prompt-dependent. Usually they require an input point, bounding box, text or mask in order to output a result. SAM and SAM2 has the option of "Segment Anything" but they perform badly on medical datasets, and specially with small structures.  
%
%These input prompts might not always be available, and second the quality of the produced mask by these models is dependent on the quality of the input prompt. 

Despite these advantages, current foundation-model-based approaches still present important limitations for lung nodule segmentation. 
First, most of them remain prompt-dependent: they require points, bounding boxes, text prompts, or previous masks to guide the segmentation. 
This is a strong assumption in clinical workflows, because such prompts are not always available and may require manual input or a separate detection model. 
Second, the quality of the final mask is highly sensitive to the quality of the prompt; small changes in the location or size of a bounding box can lead to under-segmentation, over-segmentation, or inclusion of nearby anatomical structures. 
This problem is especially relevant for pulmonary nodules, which are often small, low-contrast, irregularly shaped, and attached to vessels or pleural surfaces. 
Third, although generic models such as SAM can segment many natural image objects, their zero-shot performance is usually weaker in medical images, where CT intensity distributions, volumetric context, and subtle lesion boundaries differ substantially from the images used during pretraining. 

Therefore, in this work, we address two complementary objectives. First, we investigate the influence of prompt type and prompt quality on segmentation performance of lung nodules. 
Second, motivated by the observed robustness characteristics, we investigate whether automatically generated coarse prompts can replace manual interaction, enabling a practical prompt-free segmentation pipeline while retaining the benefits of prompt-based foundation models.
Unlike previous approaches, that use an additional network detector to produce external bounding-box prompts for SAM, the proposed model learns coarse prompt proposals in a lightweight network directly from the foundation model encoded features, making the segmentation pipeline compact and tightly coupled to the foundation model. 

\section{Method}

A common characteristic between the SAM and MedSAM models family is that they are composed by an image encoder, a prompt encoder and a mask decoder, and depending on the settings, they can produce either 1 single output mask or propose 3 output masks and their respective confidence score (estimated Intersection Over Union, IoU). 
%each of them with a confidence score. 
In the case of the second versions; SAM2~\cite{ravi2025sam} and MedSAM2~\cite{zhu2024medical} they can perform segmentation on volumetric data, which in the case of CT-scans means the whole 3D CT-volume can be used as input instead of a single 2D image slice, such as in the case of SAM and MedSAM. 

We assessed how prompt quality affects mask predictions in SAM, SAM2, MedSAM, and MedSAM2. 
%
%Each of these models also has different versions with different backbone (Vision-Transformer: ViT) sizes, and the dataset on which they were trained. In the case of SAM these correspond to ViT-b, ViT-l and ViT-h trained on SA-1B dataset. 
%
%In the case of MedSAM the checkpoints available correspond to the ViT-b backbone model trained on several medical imaging datasets, and one specific fine-tuned on FLARE22 dataset. 
%
In the baseline experiments we evaluated the predicted masks generated with each of the models on LUNA16 dataset using 3 different input prompts: point, bounding box and point$+$bounding box. The center point and the limits of the bounding boxes were extracted directly from the ground truth masks. For each model we tested the different checkpoints available in their official repositories. 

\begin{figure}
    \centering
    \includegraphics[width=0.99\linewidth]{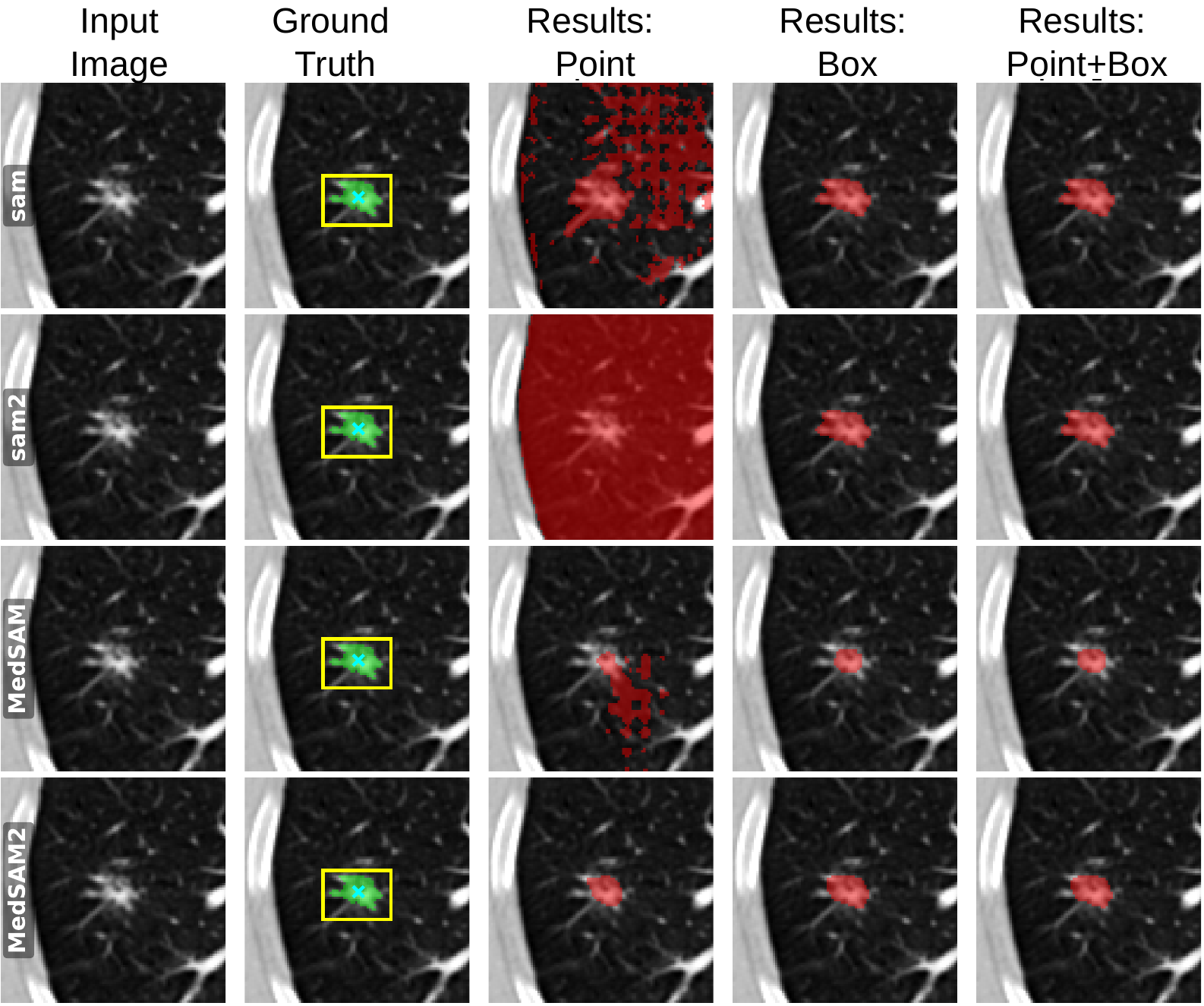}
    \caption{Samples of the predicted masks by the different foundation models. The input to the models was a 3-channel image as show in the picture, and one of the different combinations of input prompts i.e. point, box or point$+$box.}
    \label{fig:sample_predeictions}
\end{figure}

For SAM and MedSAM; as well as SAM2 and  MedSAM2 in image input modality, only nodule-containing CT slices were used, with each slice repeated three times to match the required 3-channel input. 
For SAM2 and MedSAM2 in 3D mode, the full CT volume was used.
Individual predictions were run per nodule, in the cases where $n>1$ nodules were present in the image, the final prediction was set to the combination of the $n$ predictions.
Single and multi-image outputs were evaluated for each prompt type. 
Example results are shown in Fig.~\ref{fig:sample_predeictions}.
%For SAM and MedSAM, for each CT-volume, only nodule-containing slices were used as input to the models, repeating each slice 3 times to adapt to the 3-channel input required by the models. 
% In cases where more than 1 nodule was present in the same slice, the prediction process was run separately for each of them and the final predicted mask was the combination of the $n$ slice predictions. 
%In the case of SAM2 and MedSAM2 when using the 3D data option, the whole CT-volume was used as input. 
%For all models, individual predictions were run per nodule, in the cases where $n>1$ nodules were present in the image, the final prediction was set to the combination of the $n$ predictions. 
%Single-image and multi-image output options were investigated with each of the input prompts. 
%Sample images of these experiments are shown in Fig.~\ref{fig:sample_predeictions}.

%\includegraphics[width=\linewidth]{figures/sample_output_2.pdf}
%The results of these experiments are shown in Table~\ref{tab:baseline_experiments}.
\subsection{Perturbation studies}
To test the robustness of the models towards changes in the input prompt we performed two different perturbation studies. In the first set, we tested the stability of the predictions when the input point~$(p_{x_{0}}, p_{y_{0}})$ was shifted within a random radius $r$ such that the shifted point was $(p_{x_{n}}, p_{y_{n}})=(p_{x_{0}} + r, p_{y_{0}} + r)$. 
In the cases of the bounding-box and bounding-box+point as input prompts, the center of the box was also moved to match with the shifted center point. 

The second perturbation analysis focused on changing the size of the boxes by $l$ pixels to see if adding more background information, or reduce information about the nodule boundaries would affect the performance of the models. In this case, given an initial bounding box $\{(x_{min}, y_{min}), (x_{max}, y_{max})\}$ that exactly matches the boundaries of the mask, the resized mask was $\{(x_{min} - l, y_{min} - l), (x_{max} + l, y_{max} + l)\}$ with the corresponding version of the points and boxes for the 3D input cases. 
\subsection{Supervised Decoder Adaptation for Segmentation and Synthetic Prompt Generation}
Given that one of the current limitations of image-segmentation foundation models is their reliance on input prompts, we propose adapting these models for segmentation in a fully supervised manner and prompt generation.

Considering $I \in \mathbb{R}^{Z \times H \times W}$ a CT volume, where $z$ indexes the axial slice. For each target slice $z$, the network input was defined as a three-channel 2.5D image \(x_i \in \mathbb{R}^{3 \times H \times W}\) by stacking the neighboring axial slices: $x_i =\{I_{z-1},\, I_z,\, I_{z+1}\}$.
In the fully supervised adaptation setup, the pre-trained image encoder $E_{\theta}$ was used to extract dense image embeddings $h_i$ from $x_i$ inputs. %$E_{\theta}(x_i)= h_i$ 
The resulting embeddings were passed to an FPN-style decoder \cite{lin2017feature} where multi-scale features are progressively up-sampled and combined with higher-resolution feature maps before the final mask prediction $\hat{y_i}$. 
The decoder was trained in a fully supervised manner using  annotated nodule masks.

\begin{figure}
    \centering
    \includegraphics[width=0.99\linewidth]{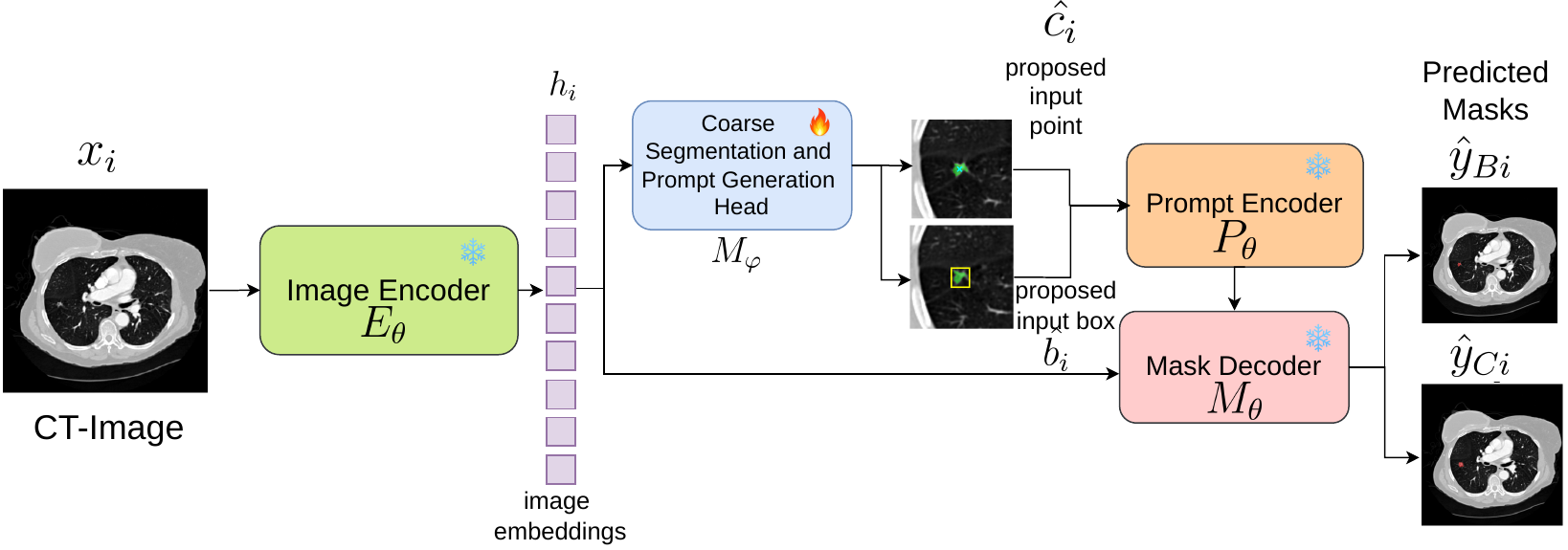}
    \caption{Prompt generation method. Foundation-model encoder embeddings are used to train the coarse segmentation and synthetic prompt-generation network. The generated points and boxes are passed to the prompt encoder to generate the final masks.}
    \label{fig:method}
\end{figure}

In the synthetic prompt-generation setup, an FPN-based decoder network $D_{\varphi}(h_i)$ is trained to predict a coarse nodule segmentation mask, and point $\hat{c}_i$ and box $\hat{b}_i$ prompts proposals from the dense feature representations $h_i$.
Point and box predictions are regressed from spatial feature maps formed by combining the refined FPN features and the predicted coarse-mask probability map. 
%The point is predicted through a heatmap-based localization head, while the bounding box is predicted using distance-map outputs that estimate the box extent around the detected nodule region.
%
The coarse mask is trained with Sørensen-Dice coefficient (DSC) and binary cross-entropy losses against the ground-truth nodule mask. The synthetic point and box predictions are supervised using normalized $L_1$ losses for point and box coordinates, together with a generalized IoU loss for the predicted bounding box.
In addition, the predicted synthetic prompts are passed through the foundation model prompt encoder $P_{\theta}(\hat{c_i}, \hat{b_i})$  and mask decoder to generate prompted-based masks $\{\hat{y}_{Ci}, \hat{y}_{Bi}\}$.
The resulting masks are compared with the ground-truth segmentation using DSC losses, encouraging the learned synthetic prompts not only to match geometric targets, but also to maximize the segmentation quality obtained from the prompted foundation model.
A schematic of the method is depicted in Fig.~\ref{fig:method}. 
%\subsection{Dataset}

We used the LUNA16 dataset, derived from the LIDC-IDRI dataset~\cite{armato2011lung}, which consists of 888 thoracic CT scans containing 1,186 annotated lung nodules, annotated by four radiologists with nodules larger than 3mm considered relevant. 

\section{Results}
%2Do: Define metrics used
The benchmark experiments were performed on the complete LUNA16 dataset. 
Performance metrics were calculated patient-wise. 
The mean DSC values were used to report the overall performance of the models in their different input prompts: point, box, point+box; and different output configuration: volume, single image and multi-image. 
%the median DSC value was reported for each of the different configurations of input prompt and types of output. 
%
For the multi-output image results, for each prompt type, results are reported for each of the 3 output channels (ch:1, ch:2, ch:3). 
The results are shown in Table~\ref{tab:baseline_experiments}. 
%
%In general, box-prompts produce better results in the single output image configuration. In the case of multi-output image, there is no consistency among the models on which channel has better results, however, there is at least always one channel that performs better.  The combination of input prompts of point+box in general does not result in better segmentations, and in some cases it slightly reduces the overall performance of the models compared with only a box as input prompt.  This might be related with the fact that the input information redundancy tends to make the model perform more conservatively, ignoring pixels that are more far away from the center point.  
%

Overall, box prompts perform better in the single-output image setting. For multi-output images, performance varies by model and channel, but at least one channel consistently performs better.
Combining point and box prompts does not generally improve segmentation and can slightly reduce performance in some cases compared with using only a box prompt.
This may be because redundant input information makes the model behave more conservative, causing it to ignore pixels farther from the center point.
The performance of point prompts is generally below a DSC of 0.1, except for MedSAM2 which achieves a performance over 0.7. 
The best results are obtained by SAM2 with a DSC of 0.89 using bounding box prompts with \emph{hierta t} and \emph{hiera s} backbones. 
Nevertheless, the results for the rest of the 3D models, i.e. SAM2 in its other configurations and MedSAM2 in all of its configurations achieve similar values of 0.88.

\begin{table}[h!]
    \centering
    \caption{\footnotesize DSC values of the different foundation models using different input prompts and different output configurations.}
    \footnotesize
    \resizebox{0.99\textwidth}{!}{%
    \begin{tabular}{c|ccc|ccc|ccc|ccc|ccc}
        \hline
        model/checkpoint
        & \multicolumn{3}{c|}{volume}
        & \multicolumn{3}{c|}{single image}
        & \multicolumn{9}{c}{multi image} \\
        \hline
        & point & box & \makecell{point \\ + box}
        & point & box & \makecell{point \\ + box}
        & \multicolumn{3}{c|}{point}
        & \multicolumn{3}{c|}{box}
        & \multicolumn{3}{c}{\makecell{point \\ + box}} \\
        \cline{8-16}
        & & &
        & & &
        
        & ch:0 & ch:1 & ch:2
        & ch:0 & ch:1 & ch:2
        & ch:0 & ch:1 & ch:2 \\
        \cline{2-16}
        \cline{2-16}
        %\hline \hline
        sam/vit\_b    & --  & -- & -- & 0.20 & 0.86 & 0.83 & 0.02 & 0.03 & 0.38 & 0.84 & 0.86 & \textbf{0.88} & 0.79 & 0.82 & \textbf{0.85} \\
        sam/vit\_l    & --  & -- & -- & 0.09 & 0.87 & 0.85 & 0.09 & 0.28 & 0.00 & 0.85 & 0.87 & 0.84 & 0.82 & 0.86 & 0.81 \\
        sam/vit\_h    & --  & -- & -- & 0.10 & 0.87  & 0.85 & 0.23 & 0.02 & 0.00 & \textbf{0.87} & 0.85 & 0.84 & \textbf{0.86} & 0.83 & 0.81 \\
        medsam/vit\_b  & --  & -- & -- & 0.10 & 0.79 & 0.79 & 0.02 & 0.04 & 0.06 & 0.76 & 0.75 & 0.75 & 0.78 & 0.77 & 0.76 \\
        medsam/flare22 & -- & -- & -- & 0.25 & 0.26 & 0.25 & 0.13 & 0.23 & 0.28 & 0.17 & 0.27 & 0.30 & 0.15 & 0.24 & 0.27\\
        sam2/hiera\_t & 0.22 & 0.33 & 0.34 & 0.13 & \textbf{0.89} & \textbf{0.89} & 0.01 & 0.32 & 0.02 & 0.46 & \textbf{0.88} & 0.83 & 0.52 & \textbf{0.88}  & 0.83 \\
        sam2/hiera\_s & 0.05 & 0.08 & 0.07 & 0.08 & \textbf{0.89} & 0.88 & 0.28 & 0.01 & 0.00 & 0.84 & 0.79 & 0.46 & 0.84 & 0.76 & 0.34 \\
        sam2/hiera\_b+ & 0.19 & 0.31 & 0.30 & 0.10 & 0.88 & 0.86 & 0.02 & 0.27 & 0.01 & 0.50 & 0.84 & 0.76 & 0.45 & 0.81 & 0.70 \\
        sam2/hiera\_l  & 0.07 & 0.14 & 0.15 & 0.09 & 0.88 & 0.87 & 0.31 & 0.02 & 0.01 & \textbf{0.87} & 0.14 & 0.76 & \textbf{0.86} & 0.09 & 0.74 \\
        medsam2/hiera\_t & \textbf{0.78} & \textbf{0.74} & \textbf{0.76} & \textbf{0.80} & 0.88 & 0.88 & \textbf{0.64} & 0.73 & \textbf{0.79} & 0.70 & 0.83 & 0.84 & 0.71 & 0.83 & 0.84 \\
        medsam2/2411 & 0.77 & 0.67 & 0.71 & 0.74 & 0.88 & 0.87 & 0.33 & 0.72 & 0.73 & 0.76 & 0.87 & 0.85 & 0.76 & 0.86 & 0.84 \\
        medsam2/CTLesion & 0.77 & 0.66 & 0.70 & 0.79 & 0.88 & 0.88 & 0.59 & \textbf{0.75} & 0.77 & 0.76 & 0.87 & 0.85 & 0.76 & 0.87 & 0.\textbf{85} \\
        
        %\hline
    \end{tabular}%
    }
    \label{tab:baseline_experiments}
\end{table}
For the perturbation studies, SAM \emph{vit\_h}, MedSAM \emph{vit\_b}, SAM2 \emph{hierta\_t} and MedSAM2 with \emph{hierta\_t} backbones were used.
In this set of experiments, single-image output and volume configurations were analyzed. 
In the center point perturbation experiments, all the 3 different input prompts were used, whereas for box-size perturbations, only box and box+point prompts were considered.
%
%The results of these experiments are shown in Fig.~\ref{fig:perturbation_results}. 

\begin{figure}
\includegraphics[width=0.99\textwidth]{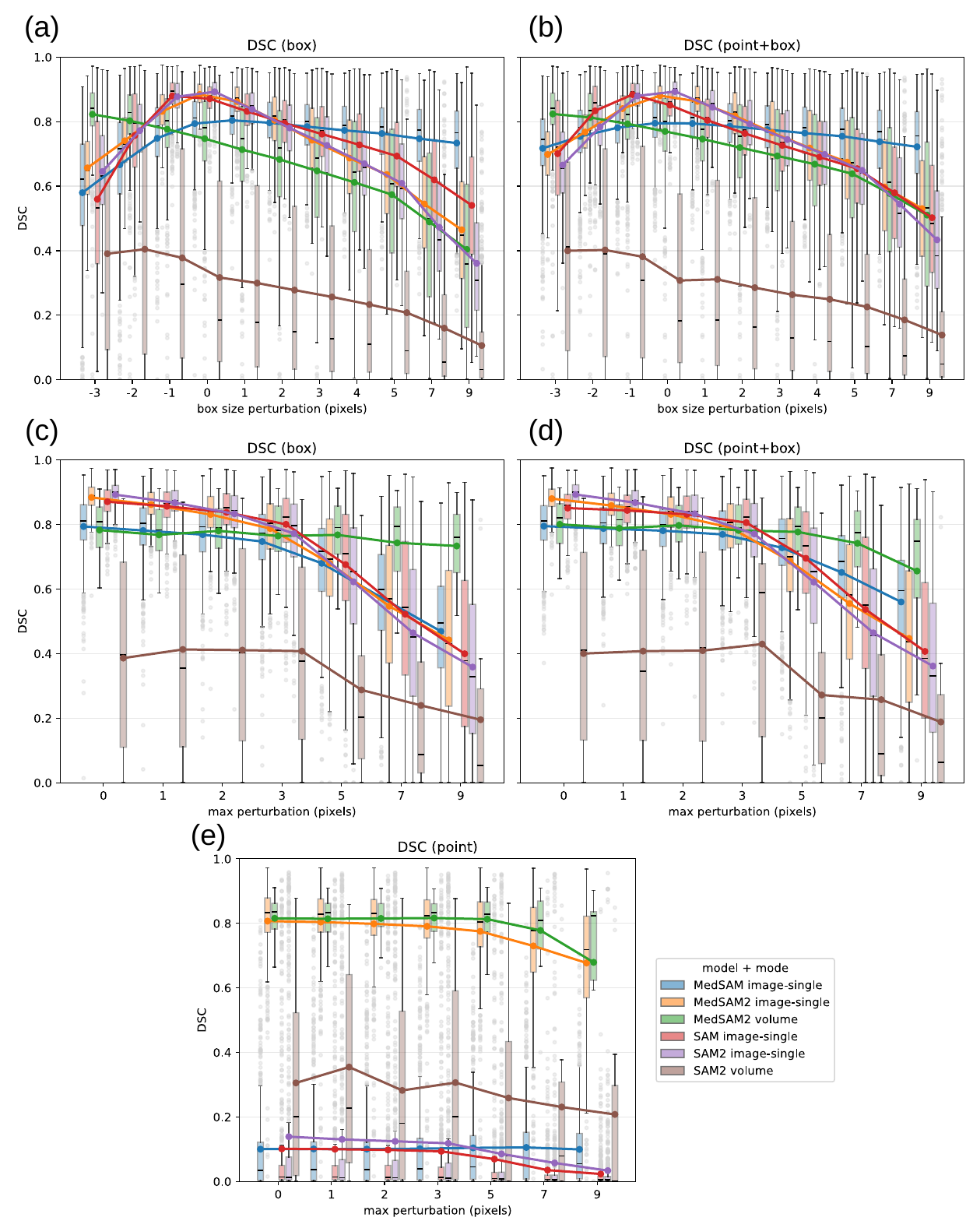}
\caption{Perturbation results. (a)-(b) Box plot comparison of the box-size perturbation experiments. (a)~using only box as input prompt, (b)~combination of point+box. (c)-(e)~box plot comparison of the center pixel perturbation experiment according for every input prompt: (c)~box, (d)~point+box, (e)~point.} 
\label{fig:perturbation_results}
\end{figure}

%Bounding box perturbations were carried out in the range $-3\leq l\leq 9$ pixels. The perturbations were performed homogeneously across all coordinates. The results of these experiments are shown in Fig.~\ref{fig:perturbation_results}(a)-(b). 
%For the single-image output configuration, the models perform worse when the box size is reduced to a size smaller than the nodule size, but the opposite behavior is observed for models with a volume output configuration. 
%In the case where the box is expanded to a size larger than the nodule, the models show in general a decline in their performance except for MedSAM which keeps mean DSC values above 0.7. 
%
Bounding boxes were uniformly perturbed across all coordinates within $-3\leq l\leq 9$ 
pixels, with results shown in Fig.~\ref{fig:perturbation_results}(a)-(b). 
In the single image output setting, reducing the box below the nodule size worsens performance, whereas volume output models show the opposite trend. 
When boxes are expanded beyond the nodule size, performance generally decreases, except for MedSAM, which maintains mean DSC values above 0.7.
%
% this might be related to the way in which MedSAM was trained.
%
The results of the center point perturbation are shown in Fig.~\ref{fig:perturbation_results}(c)-(e). Center point perturbations were carried out in a range $1\leq r\leq 9$ %$r \in [1, 9]$
pixel-radius away from the original center. 
MedSAM2 seems to be the more stable model with respect to changes in the center point, and might be related to the way in which it was trained with additional input-prompts from annotators for the same target. 
%The signal from the box is stronger than that from the point(comparison of d and e)

In the Decoder Adaptation (DA) and synthetic prompt generation experiments, the dataset was split patient-wise into training, validation, and test sets using a 0.8/0.1/0.1 ratio. The encoders and decoders of each foundation model were frozen, and only the added segmentation and prompt-proposal heads were trained.
For the proposed prompt-generation model, referred to as ``\emph{Prompt-Gen}'', the MedSAM2 encoders and decoders were used due to the stability observed in the perturbation studies. The outputs obtained from the generated point and the generated box prompts are reported separately. The proposed method was compared with U-Net and NoduleNet~\cite{tang2019nodulenet}, both of which have previously been used for lung nodule segmentation. The results of these experiments are shown in Table~\ref{tab:results_finetune}.
All models were trained for 90 epochs using AdamW optimizer with a learning rate of $1 \times 10^{-4}$. Performance was evaluated using intersection over union IoU and DSC. 
The intensities were first windowed by clipping the Hounsfield units to the range [-1000,400] HU. The clipped intensities were then linearly normalized to [0,1].
%Augmentations in the training data included random horizontal and vertical flips with probability 0.5, random affine transformations with rotations up to ±15∘, translations up to ±16 pixels, and scaling between 0.90 and 1.10. 
Intensity augmentation was applied with probability 0.8, including random brightness and contrast changes, additive Gaussian noise, and clipping back to the valid 8-bit range. 
A Gaussian blur with a 3×3 kernel was additionally applied with probability 0.1.
Overall, using SAM2 and MedSAM2 image encoders as backbones led to better results than their SAM and MedSAM counterparts. 
%whose performance was comparable to that of U-Net and NoduleNet. 
Prompt generation with the proposed model further improved the results, although the gains were modest.

The results obtained  show that foundation models such as SAM2 and MedSAM2 can achieve strong lung nodule segmentation performance when provided with good quality prompts. 
The proposed Prompt-Gen model represents a step toward reducing the need for manual prompt annotation, but its improvements remain modest and its performance is still constrained by the accuracy of the generated prompts and the limited adaptation of the frozen backbone. 
These findings suggest that prompt generation is a promising direction when using foundation models in medical image segmentation, but further work is needed to improve robustness, generalization, and end-to-end task adaptation.
Future work could incorporate explicit localization losses, uncertainty estimation, or multiple candidate prompts to improve robustness. Moreover, evaluating the method on external datasets would be important to determine whether the observed improvements generalize beyond LUNA16 and across different acquisition protocols, scanners, and annotation styles.

%The models were trained for 50 epochs using Adam with a learning rate of 1x10$^{-4}$
%Intersection over union (IoU) and DSC was used to report the performance of the models. 
%U-Net and NoduleNet~\cite{tang2019nodulenet} models were used to compare the results of the proposed method. 
%In general, using SAM2 and MedSAM2 image encoders as backbones obtains better results than its counterpart SAM and MedSAM versions, which obtain values in the range of U-Net and NoduleNet. 
%Generating prompts with the trained decoder heads helps improving the results, but these improvement are modest.   
%The models were implemented using Pytorch and trained on an Nvidia H200 GPU.  

% \begin{table}[h!]
%     \centering
%     \caption{\footnotesize Comparison of Nodule Segmentation models.}
%     \begin{tabular}{c|c|c}
%         model  & ~~~~~~DSC~~~~~~ &  ~~~~~~IoU~~~~~~ \\ \hline
%         U-Net & 0.781 & 0.640 \\
%         NoduleNet & 0.826 & 0.704 \\
%         DA - SAM & 0.692 & 0.529\\
%         DA - MedSAM & 0.683 &  0.518 \\
%         DA - SAM2  &  0.844 & 0.720 \\
%         DA - MedSAM2 &  0.826 & 0.703 \\
%         Prompt-Gen (point prompt) & 0.820 & 0.694 \\
%         Prompt-Gen (box prompt) & \textbf{0.843} & \textbf{0.728} \\
%         %\hline
%     \end{tabular}%
%     \label{tab:results_finetune}
% \end{table}
\begin{table}[h!]
    \centering
    \caption{Comparison of nodule segmentation models.}
    \label{tab:results_finetune}
    \footnotesize
    \setlength{\tabcolsep}{4pt}
    \renewcommand{\arraystretch}{0.9}
    \begin{tabular}{lcc}
        \hline
        Model & DSC & IoU \\ \hline
        U-Net              & 0.781 & 0.640 \\
        NoduleNet          & 0.826 & 0.704 \\
        DA-SAM             & 0.692 & 0.529 \\
        DA-MedSAM          & 0.683 & 0.518 \\
        DA-SAM2            & 0.844 & 0.720 \\
        DA-MedSAM2         & 0.826 & 0.703 \\
        Prompt-Gen (point) & 0.820 & 0.694 \\
        Prompt-Gen (box)   & 0.843 & 0.728 \\ \hline
    \end{tabular}
\end{table}
\section{Conclusions}
%In this work, we analyzed how the quality of input prompts affects the performance of different foundation models for medical image segmentation. We also proposed two adaptations to use these models in a prompt-less setting: first, by using the foundation models as backbones for a segmentation model, and second, by training a prompt-generation model that automatically produces input prompts for mask prediction.
In this work, we analyzed how input prompt quality affects the performance of different foundation models for medical image segmentation. We also proposed two adaptations: using foundation models encoders as frozen backbones, and training a prompt-generation model to automatically produce prompts for mask prediction.
The results of the proposed adaptation and prompt generation show that, depending on the backbone, foundation models can improve performance with respect to baseline methods in fully supervised settings. In addition, prompt generation achieved better results than using the models only as frozen backbones, although the improvement was modest. These findings suggest that prompt generation could be a valid strategy for adapting prompt-based foundation models when manual prompts are not available.
%
%The results show that, depending on the choice of backbone, foundation models can improve segmentation performance respect to baseline methods in fully-supervised set ups. Moreover, using these models to generate prompts achieved better results than using them only as frozen backbones, although the observed improvement was modest.
%These findings suggest that prompt generation is a promising strategy for adapting prompt-based foundation models to scenarios where manual prompts are not available.
%Overall, understanding what are the limitations and possible ways to adapt foundation moedls to design solutions for medical imaging segmmentation is fundamental when designing solutions for medical image segmentation based on foundation models, but also in the clinical practice if any of these models are used to generate some initial proposals of segmentation masks.
In general, understanding the limitations of foundation models and the possible strategies for adapting them is essential when designing medical image segmentation solutions. 
This is particularly relevant in clinical practice, where model sensitivity to prompt quality, prompt placement, and adaptation strategy should be carefully considered before deployment.

%In these settings, the sensitivity of the models to prompt quality, prompt placement, and model adaptation should be carefully considered before deployment.

%
% the environments 'definition', 'lemma', 'proposition', 'corollary',
% 'remark', and 'example' are defined in the LLNCS documentclass as well.
%

 %% removed for anonymized MICCAI submission.
    
    % The following acknowledgement and disclaimer sections can be removed for the double-blind review process.  If and when your paper is accepted, reinsert the acknowledgement and the disclaimer clause in your final camera-ready version.
    % IF you opted to include the acknowledgement and disclaimer sections, they will count towards the 8-page limit.

\begin{credits}
\subsubsection{\ackname} The study was supported by the Sjöberg Foundation grant 2022-489. The computations were enabled by resources provided by the National Academic Infrastructure for Supercomputing in Sweden (NAISS), partially funded by the Swedish Research Council through grant 2022-06725.
%
%\subsubsection{\discintname}
%Conflict of interests
\end{credits}

%\bibliographystyle{splncs04}
\bibliography{biblio}
%Author, F.: Article title. Journal \textbf{2}(5), 99--110 (2016)

%Author, F., Author, S.: Title of a proceedings paper. In: Editor,
%F., Editor, S. (eds.) CONFERENCE 2016, LNCS, vol. 9999, pp. 1--13.
%Springer, Heidelberg (2016). \doi{10.10007/1234567890}

%Author, F., Author, S., Author, T.: Book title. 2nd edn. Publisher,

%Author, A.-B.: Contribution title. In: 9th International Proceedings
%on Proceedings, pp. 1--2. Publisher, Location (2010)

%LNCS Homepage, \url{http://www.springer.com/lncs}, last accessed 2023/10/25
%\end{thebibliography}
\end{document}